\documentclass[letterpaper,12pt]{article}
\pdfoutput=1

\usepackage{jheppub}
\usepackage{amsmath}
\usepackage{graphicx}
\usepackage{subfig}
\usepackage{cancel}
\usepackage[dvipsnames]{xcolor}
\usepackage{color}
\usepackage{mathrsfs}
\usepackage{braket}
\usepackage{amssymb}
\usepackage{epstopdf}
\def\({\left(}
\def\){\right)}
\def\[{\left[}
\def\]{\right]}

\newpage

\title{Causal holographic information does not satisfy the linearized quantum focusing condition}

\author[a]{Zicao Fu}
\author[a]{Donald Marolf}
\author[a]{Marvin Qi}

\affiliation[a]{Department of Physics, University of California, Santa Barbara, CA 93106, USA}

\emailAdd{zicaofu@physics.ucsb.edu}
\emailAdd{marolf@physics.ucsb.edu}
\emailAdd{marvinqi@umail.ucsb.edu}

\abstract{The Hubeny-Rangamani causal holographic information (CHI) defined by a region $R$ of a holographic quantum field theory (QFT) is a modern version of the idea that the area of event horizons might be related to an entropy.  Here the event horizon lives in a dual gravitational bulk theory with Newton's constant $G_{\rm bulk}$, and the relation involves a factor of $4G_{\rm bulk}$. The fact that CHI is bounded below by the von Neumann entropy $S$ suggests that CHI is coarse-grained.  Its properties could thus differ markedly from those of $S$.  In particular, recent results imply that when $d\le 4$ holographic QFTs are perturbatively coupled to $d$-dimensional gravity, the combined system satisfies the so-called quantum focusing condition (QFC) at leading order in the new gravitational coupling $G_d$ when the QFT entropy is taken to be that of von Neumann.  However, by studying states dual to spherical bulk (anti--de Sitter) Schwarschild black holes in the conformal frame for which the boundary is a $(2+1)$-dimensional de Sitter space, we find the QFC defined by CHI is violated even when perturbing about a Killing horizon and using a single null congruence. Since it is known that a generalized second law (GSL) holds in this context, our work demonstrates that the QFC is not required in order for an entropy, or an entropy-like quantity, to satisfy such a GSL.}

\begin{document}
\maketitle

\section{Introduction}
\label{sec:intro}

The Hubeny-Rangamani causal holographic information (CHI) \cite{Hubeny:2012wa} is a modern version of the idea that event horizons carry an entropy $A/4G_{\rm bulk}$. For any region $R$ in a holographic quantum field theory (QFT), the domain of dependence $D(R)$ in the QFT defines future and past sets $I^\pm[D(R)]$ in the bulk gravitational dual with associated past and future bulk horizons $\partial {I}^\pm[D(R)]$.  Taking the causal surface $C(R)$ to be the intersection
\begin{equation}
\label{eq:CHI}
C(R) :=\partial {I}^+[D(R)]\cap \partial {I}^-[D(R)],
\end{equation}
with area $A[C(R)]$,  CHI$(R)$ is $A[C(R)]/(4G_{\rm bulk})$.  We will explore whether CHI satisfies a particular condition known as the linearized quantum focusing condition, about which we will say more below.

CHI is already known to satisfy several other interesting properties.  First,
CHI$(R)$ is bounded below by the Hubeny-Rangamani-Takayanagi (HRT) entropy $S_{\rm HRT}(R)$ \cite{Hubeny:2007xt,Wall:2012uf}. As a result, it has been proposed (see e.g. Ref. \cite{Kelly:2013aja}, though see also Ref. \cite{Engelhardt:2017wgc}) that CHI might quantify some coarse-grained entropy in the holographic QFT.  Second, though CHI is generally infinitely greater than $S_{\rm HRT}$ \cite{Freivogel:2013zta}, when the intersection of $\partial {I}^+(R)$ or $\partial {I}^-(R)$ with the asymptotically locally anti--de Sitter (AlAdS) boundary is a Killing horizon, it was shown in Ref. \cite{Bunting:2015sfa} that\footnote{At least in the so-called universal sector, dual to pure Einstein-Hilbert gravity in the bulk.} CHI can be at most finitely greater.

Subject to the same footnote, Ref. \cite{Bunting:2015sfa} also explored the coupling of the $d$-dimensional holographic QFT to Einstein-Hilbert gravity via some small $d$-dimensional Newton constant $G_d$.  (Note that this $G_d$ has nothing to do with the Newton constant $G_{\rm bulk}$ that controls the bulk dual.)
In classical general relativity the second law of black hole mechanics states that the area of black holes cannot decrease.   But in the context of semiclassical gravity, Bekenstein \cite{Bekenstein:1973ur} proposed it to be replaced by the generalized second law (GSL), requiring the non-decrease of the generalized entropy.  This quantity is the sum
\begin{equation}
S_{\rm gen}=S_{\rm BH}+S_{\rm QFT}
\end{equation}
of black hole entropy (taken to be $S_{\rm BH}=A/(4G_d)$) and the entropy of quantum fields outside. $S_{\rm gen}$ can be evaluated on codimension-2 surfaces that cut through every generator of the horizon, and the GSL requires $S_{\rm gen}$ to be non-decreasing when the cut is deformed toward the future. At least to first order in $G_d$ in perturbation theory about Killing horizons, Ref. \cite{Bunting:2015sfa} showed the GSL to hold when one takes $S_{\rm QFT}$ to be CHI.

Now, the GSL and related ideas involving gravity and entropy were used in Ref. \cite{Bousso:2015mna} to motivate a so-called quantum focusing condition (QFC) for semi-classical theories of gravity.  This condition again involves the generalized entropy $S_{\rm gen}$, and in particular considers second variations defined by a pair of (say, outgoing) null congruences orthogonal to a (now arbitrary) codimension-2 surface $\sigma $.   We shall test this condition at first order in $G_d$ for  perturbations about Killing horizons, in which case the constancy along each generator of the area element $\sqrt{h}$ defined by the cut allows the first-order QFC to be written as
\begin{equation}
\label{eq:QFC}
0 \ge \frac{d}{d \lambda_1} \frac{d}{d \lambda_2} S_{\rm gen} = \frac{d}{d \lambda_1} \frac{d}{d \lambda_2} S_{\rm QFT} - \int_\sigma d^{d-2}y\sqrt{h} 2\pi T_{\alpha \beta} k_1^\alpha k_2^\beta,
\end{equation}
in terms of the stress tensor $T_{\alpha \beta}$ of the holographic quantum field theory (i.e., in terms of the boundary stress tensor of the bulk dual). In Eq. \eqref{eq:QFC}, $\lambda_{1}$ and $\lambda_{2}$ are affine parameters along the Killing horizon associated with null generators $k_{1}^{\alpha}\partial _{\alpha }=d/(d\lambda_{1})$ and $k_{2}^{\alpha}\partial _{\alpha }=d/(d\lambda_{2})$, and $y^i$ for $i = 1,\cdots , d-2$ are coordinates on the codimension-2 surface $\sigma $. The derivatives act on $S_{\rm gen}$ by moving the cut on which it is evaluated.  The equality in Eq. \eqref{eq:QFC} uses the Raychaudhuri equation and the gravitational equation of motion at first order in $G_d$, which together relate derivatives of $S_{\rm BH}=A/(4G_d)$ to the flux of QFT energy-stress tensor $T_{\alpha \beta}$ across the horizon.

We will further specialize below to the ``single-flow'' case\footnote{It would be natural to call this the "diagonal" case.  But that term was used in Ref. \cite{Bousso:2015mna} to refer to those single-flow cases where $k = d/d\lambda$ has delta-function-like support on a single generator of the congruence; i.e., its use of the term ``diagonal" implicitly implied ``diagonal in a local basis."  We thus use ``single-flow" to avoid confusion.} $\lambda_1=\lambda_2=\lambda$, which is most closely related to the GSL.   But the QFC in principle allows $\lambda_{1}$ and $\lambda_{2}$ to be associated with distinct $k_{1}$ and $k_{2}$ respectively.  Of course, since both sets of generators must be outgoing, wherever $k_2^\alpha$ is non-vanishing we must have $k_1^\alpha = f(g) k_2^\alpha$ for some function $f$ of the null horizon generators $g$.  However, this nevertheless allows one to discuss separately the ``off-diagonal'' contributions where the supports of $k_{1}$ and $k_{2}$ do not overlap and the single-flow contributions to Eq. \eqref{eq:QFC} where $k_{1} = k_2$.  As described in Ref. \cite{Bousso:2015mna}, the off-diagonal QFC is directly related to strong subadditivity (SSA) of $S_{\rm QFT}$.  In contrast, the single-flow terms where $k_{1}$ and $k_{2}$ are supported on a single generator lead to the so-called quantum null energy condition (QNEC).\footnote{Recently, the fine-grained QNEC has been proved for free QFTs on Killing horizons \cite{Bousso:2015wca}, for QFTs in flat spacetime that flow to non-trivial conformal field theories in the ultraviolet \cite{Balakrishnan:2017bjg}, and for holographic QFTs in $d\le 5$ dimensions under certain circumstances \cite{Fu:2017evt, Akers:2017ttv}.}

Now, while SSA is a well-known property of the von Neumann entropy, this property does {\it not} hold for CHI \cite{Hubeny:2012wa}.  One would thus expect the off-diagonal QFC to fail as well.  But it is less clear what to expect in the single-flow case which would be used to prove the GSL.  We therefore test the single-flow case below, using CFT\footnote{Throughout the text, we adopt standard acronyms including CFT (conformal field theory), dS (de Sitter), and AdS (anti--de Sitter).} states on 2+1 dS space whose bulk dual is described by global AdS-Schwarzschild black holes.  In contrast to an analogous successful test of the $d=2$ CHI QNEC in Ref. \cite{Fu:2016avb}, we find that the single-flow linearized CHI QFC fails for $d=3$.  In particular, a violation occurs in a time interval that includes the moment when the bulk causal surface $C(R)$ changes topology.  We work at leading order in $G_d$ and perturb about a Killing horizon.  Since this is precisely the context where Ref. \cite{Bunting:2015sfa} showed a CHI GSL to hold, our work demonstrates that the QFC is not required in order for an entropy-like quantity to satisfy such a GSL at this order in $G_d$.

As a useful aside, we comment that the success or failure of the linearized single-flow CHI QFC is intimately connected to the behavior of the bulk horizon area at caustics.  To see this connection, we remind the reader that from Ref. \cite{Bunting:2015sfa} we know that changes in CHI along a boundary horizon can be separated into two parts as follows:
\begin{equation}
\frac{d}{d\lambda }S_{\rm CHI}=\frac{1}{4G_{\rm bulk}}\left(\frac{d}{d\lambda }A_{\rm bulk}+F\right),
\end{equation}
where $(dA_{\rm bulk})/(d\lambda )$ is the change in area of the associated bulk horizon when a bulk cut is displaced toward the future along the generators and the remainder $F$ is the flux of such generators through a cut-off surface near the AlAdS boundary.   The connection to caustics comes through the further observation of Ref. \cite{Bunting:2015sfa} that on boundary Killing horizons we have
\begin{equation}
\frac{1}{4G_{\rm bulk}}\frac{d}{d\lambda }F=\int _{\sigma }d^{d-2}y\sqrt{h}2\pi T_{\alpha \beta }k^{\alpha }k^{\beta }.
\end{equation}
Thus, CHI satisfies Eq. \eqref{eq:QFC} if and only if
\begin{equation}
\label{eq:focusing}
\frac{d^2}{d\lambda ^2}A_{\rm bulk}\le 0.
\end{equation}
When there is no caustic on the bulk horizon, the bulk focusing theorem (a direct consequence of the Raychaudhuri equation) guarantees that Eq. \eqref{eq:focusing} holds. Failures of Eq. \eqref{eq:QFC} can thus arise only from the behavior of $A_{\rm bulk}$ at caustics.  Said differently, if Eq. \eqref{eq:QFC} were to hold generally, it would imply a surprising constraint on the effect of caustics on areas of bulk horizon slices.

We begin below in section \ref{sec:spacetime} with some preliminary comments on our setting the propagation of null geodesics in AdS$_4$-Schwarzschild, and the associated effect on CHI.  The direct test of the linearized single-flow QFC is then performed in section \ref{sec:testing}.  We close with some final discussion in section \ref{sec:discussion}.

\section{Setting the stage}
\label{sec:spacetime}

We consider $d=3$ CFT states dual to the AdS$_4$-Schwarzschild spacetime
\begin{equation}
ds^2 = -\left(1-\frac{\mu}{r} + \frac{r^2}{\ell ^2}\right) dt^2 + \frac{dr^2}{1-\frac{\mu}{r} + \frac{r^2}{\ell ^2}} + r^2 \left(d\theta ^2+\sin ^2\theta d\phi ^2\right),
\end{equation}
where $\ell $ is the AdS length scale. Introducing a new coordinate $z:=\ell ^2/r$, the line element becomes
\begin{equation}
ds^2 = \frac{\ell ^2}{z^2}\left[-f\left(z\right) dt^2 + \frac{dz^2}{f\left(z\right)} + \ell ^2 \left(d\theta ^2+\sin ^2\theta d\phi ^2\right) \right],
\end{equation}
where $f\left(z\right)\equiv 1 - (\mu z^3)/\ell ^4 + z^2/\ell ^2$. The boundary of this spacetime (located at $z = 0$) is an Einstein static universe (ESU) $\mathbb{R} \times S^2$, with the metric
\begin{equation}
ds^2_{\rm ESU}=-dt^2+\ell ^2\left(d\theta ^2+\sin ^2\theta d\phi ^2\right).
\end{equation}
However, by a change of conformal frame we may instead take the boundary metric to be
\begin{equation}
\label{eq:frame}
ds^2_{\rm dS}=\Omega ^2ds^2_{\rm ESU},
\end{equation}
where $\Omega =-1/[\sin (t/\ell )]$. We shall use this representation below.   In particular, the future and past dS boundaries occur at $t=-\pi \ell$ and $t=0$.

For a given region $R $ on a boundary Cauchy surface, the causal surface $C(R)$ in the bulk is given by Eq. \eqref{eq:CHI}. In general, for $d\ge 3$, the shape of $C(R)$ can be complicated. But we consider here only cases where $D(R)$ consists of points in $\rm dS_3$ to the past of some point $p^+$ that are also to the future of some point $p^-$, so that the bulk pasts and futures $I^\pm[D(R)]$ are just $I^-(p^+)$ and $I^+(p^-)$.  As a result, Eq. \eqref{eq:CHI} becomes
\begin{equation}
C(R)=\partial I^-[p^+]\cap \partial I^+[p^-].
\end{equation}

Below, we take $p^+ = (t_+,\theta_+) = (0,0)$ and $p^- = (t_-,\theta _-) = (t_-,0)$ with $t_- <0$, so that we may choose $R$ to be the interval $t_-/(2\ell )<\theta <-t_-/(2\ell )$ at $t = t_-/2$.   This case is not generic, as $p^+$ and $p^-$ are related by an ESU time-translation.  But we will see that it suffices to show the counterexample mentioned above.  Since $p^+$ lies on the future dS boundary, its past light cone is a dS Killing horizon $\mathcal{H_{\rm bndy}}$.

Before performing our test, we must address the fact that CHI is infinite because the causal surface extends to the boundary.   In doing so, we note that (without renormalization), the holographic stress tensor $T_{\alpha \beta}$ also diverges.  Indeed, when $R$ ends on a boundary Killing horizon like ${\cal H}_{\rm bndy}$ and $k_1=k_2$, the results of Ref. \cite{Bunting:2015sfa} imply that these divergences cancel and the right-hand side of Eq. \eqref{eq:QFC} is in fact finite. To be specific, in this context Ref. \cite{Bunting:2015sfa} showed that CHI may be rendered finite by  using a Fefferman-Graham regulator and the same counterterms as for the HRT entropy.  But it is also known \cite{Bousso:2015mna} that the HRT counterterm cancels that required to renormalize the stress tensor in this context.  Thus the right-hand side of Eq. \eqref{eq:QFC} is finite, and may be computed by separately renormalizing each term in this way.

This property makes the dS conformal frame useful conceptually.  But in practice it is convenient to work in the ESU conformal frame.  We thus note that, since $d=3$ requires only a single entropy counterterm proportional to the area of $\partial R$, there is no conformal anomaly and the renormalized CHI entropy in the dS conformal frame is exactly the same as that in the original ESU conformal frame; i.e.,
\begin{equation}
\label{eq:ren}
S_{\text{CHI, ren}}=\frac{1}{4G_{\rm bulk}} \underset{{{z}_{0}}\to 0}{\mathop{\lim }}\, \left[\text{Area}_{d=2}\left(C_{z>z_0}[R ]\right)-\frac{\ell ^2}{z_0}\text{Area}_{d=1}\left(\mathcal{H_{\rm bndy}}\right)\right].
\end{equation}
Indeed, we could compute the entire quantity on the right-hand side of Eq. \eqref{eq:QFC} in the ESU conformal frame so long as we use parameters $\lambda_1=\lambda_2$ that are affine with respect to the dS conformal frame.\footnote{Failing to do so would introduce extra terms related to the expansion $\vartheta $ and its derivative $\dot{\vartheta }$ of ${\cal H}_{\rm bndy}$ in the ESU conformal frame so that the equality in Eq. \eqref{eq:QFC} would no longer hold.  Thus we can no longer use the right-hand side of Eq. \eqref{eq:QFC} to argue that the result is identical in the two conformal frames and, indeed, we would not expect it to be so.}  But rather than keep track of this last restriction, we instead simply use the ESU conformal frame as an intermediate step in computing the right-hand side of Eq. \eqref{eq:QFC} as defined by the dS conformal frame.

To study Eq. \eqref{eq:QFC} we must first locate the past bulk light cone of $p^+ =(t=0,\theta =0,z=0)$. This is a bulk Rindler horizon from the point of view of the ESU conformal frame.   Since $p^+$ is at the north pole of the sphere, the azimuthal coordinate $\phi$ is undefined at $p^+$.   But the corresponding rotational symmetry means that each null generator of the horizon has a fixed value of $\phi$.   The remaining equations for these geodesics are given by energy conservation, angular momentum conservation, and the null condition:
\begin{align}
E&=\frac{\ell ^2}{z^2}f\left(z\right)\frac{dt}{d\bar{\lambda }},\\
L&=\frac{\ell ^4}{z^2}\frac{d\theta }{d\bar{\lambda }},\\
0&=-f\left(z\right)\left(\frac{dt}{d\bar{\lambda }}\right)^2+\frac{1}{f\left(z\right)}\left(\frac{dz}{d\bar{\lambda }}\right)^2+\ell ^2\left(\frac{d\theta }{d\bar{\lambda }}\right)^2,
\end{align}
where $\bar{\lambda }$ is a null affine parameter for each geodesic. Defining a dimensionless parameter $\eta :=L/(\ell E)$, the above equations become
\begin{align}
\label{eq:thetaoft}
\frac{d\theta }{dt}&=\frac{\eta }{\ell }f\left(z\right),\\
\label{eq:zoft}
\left(\frac{dz}{dt}\right)^2&=f^2\left(z\right)-\eta ^2f^3\left(z\right).
\end{align}
Generally, $z$ is not a monotonic function of $t$. To avoid the associated sign ambiguity, we differentiate Eq. \eqref{eq:zoft} once more to obtain
\begin{equation}
\frac{d^2z}{dt^2}=\left[f\left(z\right)-\frac{3}{2}\eta ^2f^2\left(z\right)\right]\frac{df\left(z\right)}{dz}.
\end{equation}
Numerically solving this equation with initial conditions ${{\left. z \right|}_{t=0}}=0$ and ${{\left. [(dz)/(dt)] \right|}_{t=0}}=-\sqrt{1-{{\eta }^{2}}}$ yields $z=z\left(t\right)$, from which one may further numerically integrate Eq. \eqref{eq:thetaoft} to find $\theta \left(t\right)$.

This information can be used to compute CHI.  In particular, we consider the family of boundary regions $R$ defined above, such that at each $t \in (-\pi \ell ,0)$ we have $R$ to be $t/\ell <\theta <-t/\ell $.   By time-reflection symmetry, the causal surface $C(R)$ also occurs at the same time $t$. It is just the intersection of the bulk past light cone $\partial I^-(p^+)$ and with the surface of constant time coordinate $t$.   For simplicity, we refer to this surface as $C(t)$ below.

The other ingredient in Eq. \eqref{eq:QFC} is the stress tensor, which for our bulk spacetime takes the form \cite{Myers:1999psa}
\begin{equation}
T_{\alpha \beta }=\frac{\mu }{16\pi G_{\rm bulk}\ell ^2}\left(3\delta ^0_{\alpha }\delta ^0_{\beta }+g_{\alpha \beta }\right),
\end{equation}
where $g_{\alpha \beta }$ denotes the boundary metric. Since the stress tensor is defined as $T_{\alpha \beta }:=[-2/(\sqrt{|g|})](\delta S)/(\delta g^{\alpha \beta })$, a Weyl rescaling $\tilde{g}_{\alpha \beta }=\Omega ^2g_{\alpha \beta }$ of the boundary metric yields $\tilde{T}_{\alpha \beta }=\Omega ^{-d+2}T_{\alpha \beta }$. Combining this with the fact that the dS affine parameters are related by $d/(d \lambda )=\Omega ^{-2}[d/(dt)]$, we obtain
\begin{equation}
\left(T_{\alpha \beta }k^{\alpha }k^{\beta }\right)_{\rm dS}=\left(T_{\alpha \beta }k^{\alpha }k^{\beta }\right)_{\rm ESU}\Omega ^{-5}.
\end{equation}

Finally note that, on the Killing horizon where we evaluate the right-hand side of Eq. \eqref{eq:QFC}, since $\theta=-t/\ell $ the volume element on any cut satisfies $\sqrt{h} dy= \Omega \sin \theta \ell d\phi =\ell d\phi $.  To test our first-order single-flow QFC (Eq. \eqref{eq:QFC}), we thus need only check positivity of the quantity
\begin{equation}
\label{eq:QNECforCHI}
Q\equiv -\frac{1}{2\pi }\frac{d^2S_{\rm gen}}{d\lambda^2} = \frac{3\mu }{8G_{\rm bulk}\ell }\sin ^5\left(\frac{t}{\ell }\right)-\frac{1}{2\pi }\sin ^2\left(\frac{t}{\ell }\right)\frac{d}{dt}\left[\sin ^2\left(\frac{t}{\ell }\right)\frac{dS_{\text{CHI, ren}}\left(t\right)}{dt}\right].
\end{equation}

\begin{figure}
	\centering
	\includegraphics[width=0.5\textwidth]{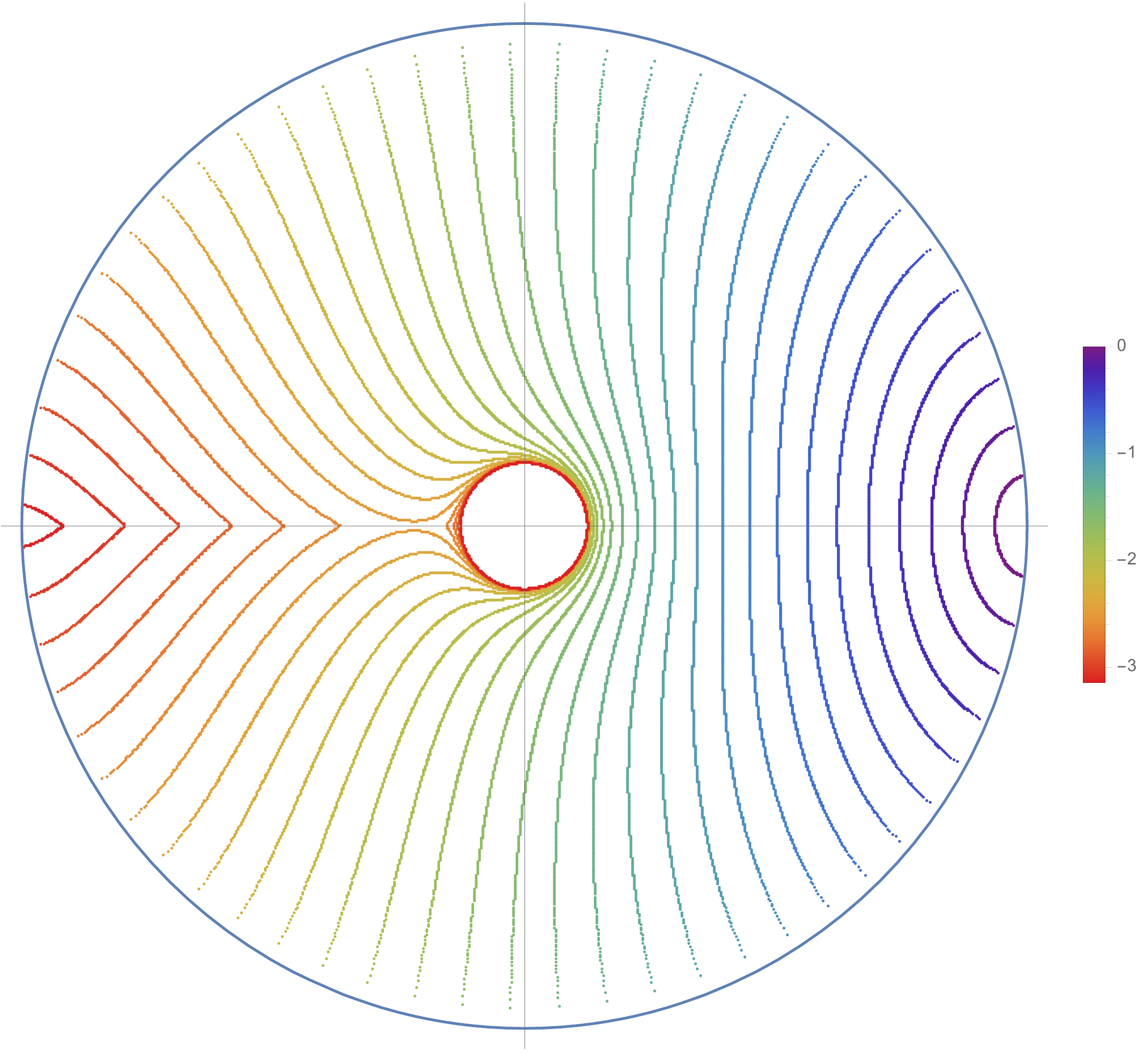}
	\caption{Our surfaces $C(R)$ projected onto a common constant-time hypersurface and drawn by taking standard polar coordinates $\rho, \theta$ on the plane to represent $\rho \equiv \arctan (r/\ell )$ and $\theta$ in AdS$_4$-Schwarzschild. The azimuthal $\phi$ is suppressed.   The outer circle denotes the boundary, the inner circle is the horizon, and the point $p_+$ is located on the right hand side of the diagram.  The surfaces do not penetrate the interior of the black hole. At some finite time $t=t_*$, the causal surface forms a cusp and disconnects. In the figure we have chosen $r_{\rm EH}/\ell =1/5$, for which $t_*\approx -2.5$.}
	\label{fig:CHIsurfaces}
\end{figure}

\section{Testing the CHI QFC}
\label{sec:testing}

To compute $Q$ in Eq. \eqref{eq:QNECforCHI} we now locate the causal surfaces $C(t)$ and compute their area. As noted above, $C(t)$ is just the cut of the bulk past light cone $\partial I^-(p^+)$ at time coordinate $t$. Each $C(t)$ is thus a codimension-2 surface described by $t=\text{const}$, $\theta =\theta \left( z \right)$, where $z$ ranges from $0$ to some maximal value $z_{\max }$.  The projections of such cuts onto a common constant-$t$ surface are shown in Fig. \ref{fig:CHIsurfaces}. The metric induced on a causal surface is
\begin{equation}
ds^2=\frac{\ell ^2}{z^2}\left[\left(\frac{1}{f\left(z\right)}+\ell ^2\left(\frac{d\theta \left(z\right)}{dz}\right)^2\right)dz^2+\ell ^2\sin ^2\left(\theta \left(z\right)\right)d\phi ^2\right],
\end{equation}
and the area of each causal surfaces is given by
\begin{equation}
\text{Area}_{d=2}\left(C_{z>z_0}[R ]\right)=2\pi \int_{z_0}^{z_{\max }}\frac{\ell ^3\sin \left(\theta \left(z\right)\right)}{z^2}\sqrt{\frac{1}{f\left(z\right)}+\ell ^2\left(\frac{d\theta \left(z\right)}{dz}\right)^2}dz.
\end{equation}

\begin{figure}
	\centering
	\includegraphics[width=0.5\textwidth]{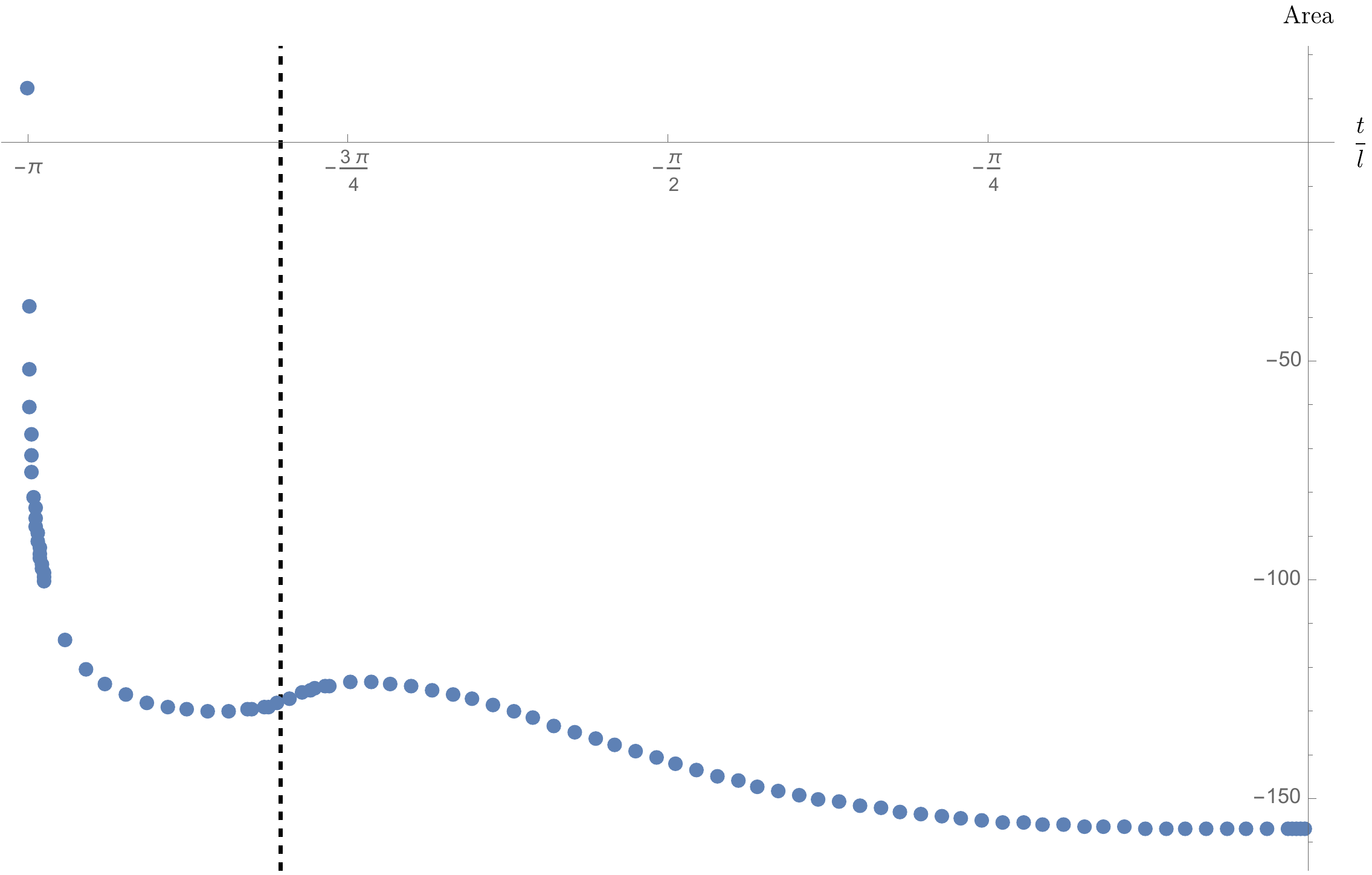}
	\caption{The renormalized area as a function of time $t$. Both in the far past and in the far future, the renormalized area is a decreasing function of time. In a small neighborhood of the moment $t_*$ when the topology of the CHI surface changes, the renormalized area increases with time. In the figure we have chosen $r_{\rm EH}/\ell =1/5$, for which $t_*\approx-2.5$ and is indicated by the dashed vertical line. }
	\label{fig:areas}
\end{figure}

The renormalized area $A_{\rm ren}=4G_{\rm bulk}S_{\text{CHI, ren}}$ is shown in Fig. \ref{fig:areas} as a function of the time $t$. Note that $(dA_{\rm ren})/(dt)$ changes sign twice. Both in the far past and in the far future, the renormalized area is a decreasing function of time. But the renormalized area increases with time in a small neighborhood of the time $t_*$ when the topology of $C$ changes.

\begin{figure}
	\centering	\includegraphics[width=0.45\textwidth]{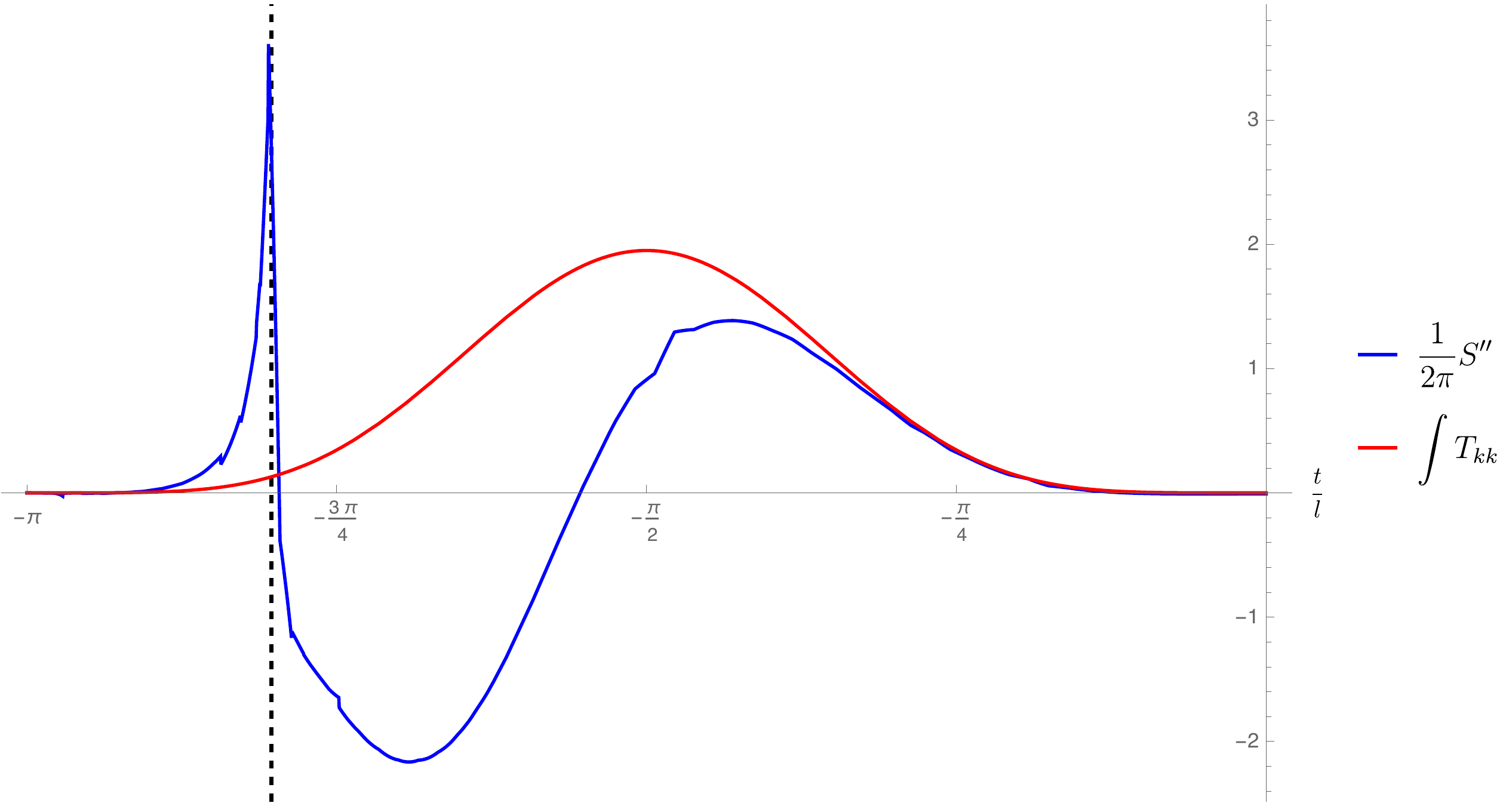}	\includegraphics[width=0.45\textwidth]{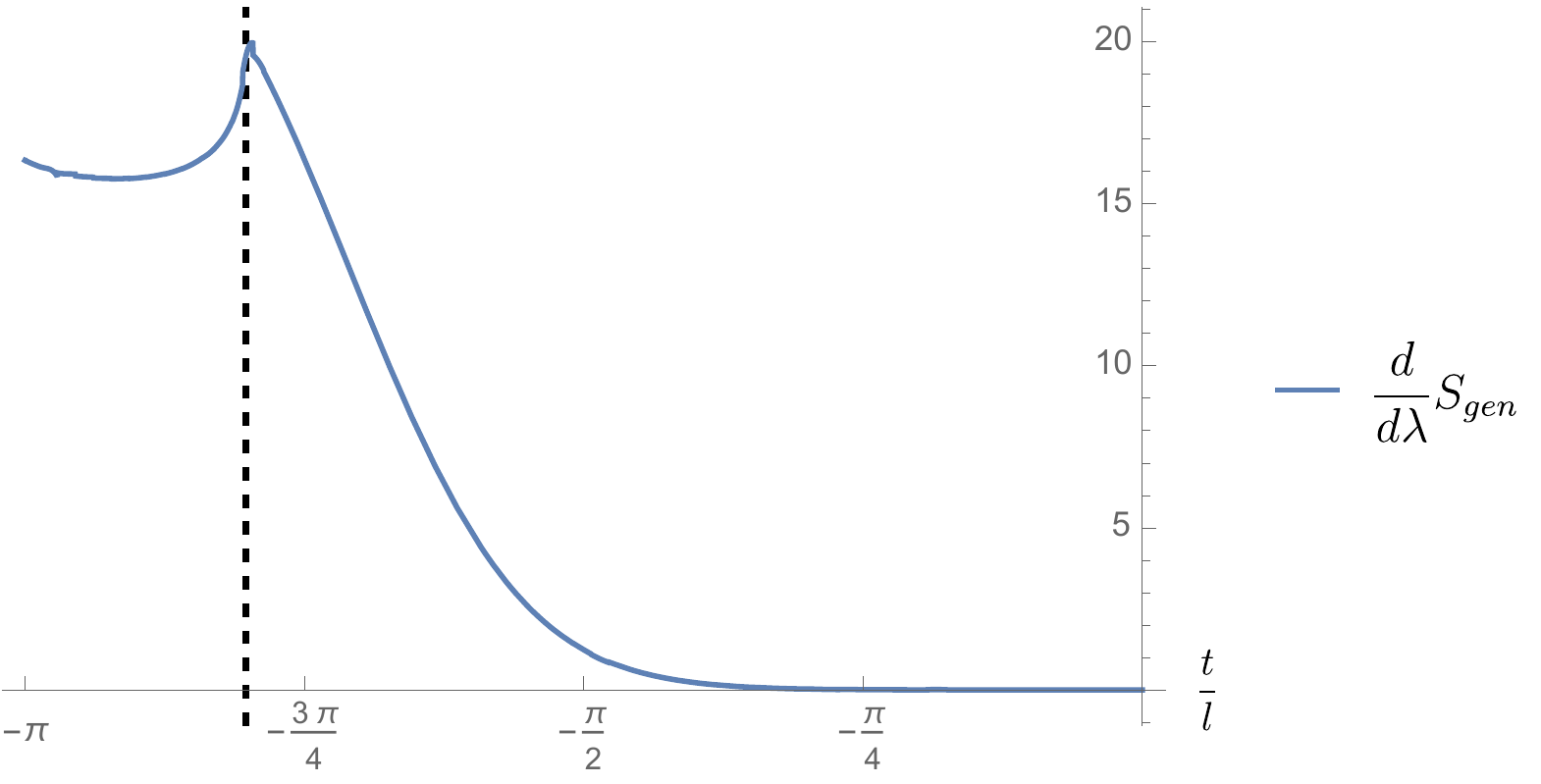}
	\caption{\textbf{Left:} Our test of the linearized QFC.  Note that $S''$ diverges when the topology of the causal surface changes. The time $t=t_*$ of the transition  is indicated by the vertical dashed line. The linearized QFC is violated in the neighborhood of this phase transition where $S''/(2\pi )$ exceeds $\int dy\sqrt{h} T_{kk}$. In the figure we have chosen $r_{\rm EH}/\ell =1/5$, for which $t_* \approx -2.5$. \textbf{Right:} The corresponding test of the GSL shows that the GSL is not violated.}
	\label{fig:qneccheck}
\end{figure}

We may now compute $Q$ in Eq. \eqref{eq:QNECforCHI}.  Each of the terms in Eq. \eqref{eq:QNECforCHI} is plotted separately in Fig. \ref{fig:qneccheck} (left). As one can see, the $T_{kk}$ term is smooth but the $S''$ term diverges (to positive infinity) at $t_*$. Thus  $S''/(2\pi )$ near $t_*$ fails to be bounded above by $\int dy\sqrt{h} T_{kk}$ and Eq. \eqref{eq:QFC} is violated.

As a check on our results, recall that Ref. \cite{Bunting:2015sfa} showed the GSL defined by CHI to be satisfied at this order in $G_3$ whenever $R$ ends on a Killing horizon (such as ${\cal H}_{\rm bndy}$).   Thus $(dS_{\rm gen})/(d\lambda ) \ge 0$. Our example respects this result, as many be seen by noting (see Fig. \ref{fig:areas}) that $S_{\rm gen}$ is finite at $t=0$ ($\lambda = \infty$), so that $(dS_{\rm gen})/(d\lambda )\rightarrow 0$ as $\lambda \rightarrow \infty$.  One may therefore integrate $Q$ to obtain
\begin{equation}
\frac{dS_{\rm gen}}{d\lambda }=2\pi \int_{\lambda }^{+\infty }Qd\lambda .
\end{equation}
Changing the time parameter from $\lambda $ to $t$, this becomes
\begin{equation}
\label{eq:GSLtesting}
\frac{dS_{\rm gen}}{d\lambda }=2\pi \int_{t}^{0}Q\left(t\right)\frac{1}{\sin ^2\frac{t}{\ell }}dt.
\end{equation}
Plotting this quantity in Fig. \ref{fig:qneccheck} (right), we find that it is always positive. Thus our example shows that the linearized (single-flow) QFC is not required for an entropy to satisfy the (linearized) GSL.

\section{Discussion}
\label{sec:discussion}

Our work above found that the linearized single-flow QFC defined by CHI is violated for $d=3$ holographic CFTs states on a dS background that are dual to global AdS$_4$-Schwarzschild black hole spacetimes. Here the adjective ``linearized'' refers to the term of order zero in the coupling $G_3$ of our $d=3$ holographic CFT to $d=3$ Einstein-Hilbert gravity.  (The term of order $G_3^{-1}$ vanishes because we evaluate the GSL on a Killing horizon of the $\rm dS_3$ background geometry.)  The violation occurs near the point at which the bulk causal surface changes topology and is associated with the formation of bulk caustics.
In contrast, an otherwise similar test was performed in Ref. \cite{Fu:2016avb} for $d=2$ in which the topology of $C(R)$ did not change and no violation was observed.  

Despite our violation of the CHI QFC, the linearized GSL defined by CHI holds in our example.  We thus conclude that an entropy, or an entropy-like quantity, need not necessarily respect a QFC in order to satisfy a GSL at this order in $G_3$. One would, of course, like to better understand just what CHI represents in the dual QFT and, especially in light of Ref. \cite{Engelhardt:2017wgc}, the extent to which it acts like an entropy.  While the first order GSL result of Ref. \cite{Bunting:2015sfa} is non-trivial, it remains to see what other useful properties CHI might satisfy.  It would also be interesting to investigate the QFCs defined by other entropy-like quantities associated regions of holographic QFTs, and in particular for those defined in Ref. \cite{Grado-White:2017nhs} by the area of marginally trapped surfaces in the dual bulk spacetime. Since these latter quantities are analogous to that discussed in Ref. \cite{Engelhardt:2017aux} for the global QFT, we may expect the arguments of Ref. \cite{Engelhardt:2017aux} to apply to them as well.  If so, then in contrast to CHI, such quantities would indeed represent coarse-grained entropies for the dual QFT.

\section*{Acknowledgements}
This work was supported in part by the Simons Foundation and by funds from the University of California.

\bibliographystyle{utcaps}
	\cleardoublepage
\phantomsection
\renewcommand*{\bibname}{References}

\bibliography{all}

\end{document}